# Mechanically tunable spontaneous vertical charge redistribution in few-layer WTe$_2$

Zeyuan Ni[1,*], Emi Minamitani[1,*], Kazuaki Kawahara[2], Ryuichi Arafune[3], Chun-Liang Lin[2], Noriaki Takagi[2], Satoshi Watanabe[1,*]

[1]Department of Materials Engineering, The University of Tokyo, Tokyo 113-8656, Japan

[2]Department of Advanced Materials Science, The University of Tokyo, 5-1-5 Kashiwanoha, Kashiwa, Chiba 277-8561, Japan

[3]International Center for Materials Nanoarchitectonics (WPI-MANA), National Institute for Materials Science, 1-1 Namiki, Tsukuba, Ibaraki 305-0044, Japan

*Corresponding author: zeyuan_ni@cello.t.u-tokyo.ac.jp; eminamitani@cello.t.u-tokyo.ac.jp; watanabe@cello.t.u-tokyo.ac.jp

**Abstract**

Broken symmetry is the essence of exotic properties in condensed matters. WTe$_2$ exceptionally takes a non-centrosymmetric crystal structure in the family of transition metal dichalcogenides, and exhibits novel properties[1–4], such as the nonsaturating magnetoresistance[1] and ferroelectric-like behavior[4]. Herein, using the first-principles calculation, we show that unique layer stacking in WTe$_2$ generates surface dipoles with different strengths on the top and bottom surfaces in few-layer WTe$_2$. This leads to a layer-dependence for electron/hole carrier ratio and the carrier compensation responsible for the unusual magnetoresistance. The surface dipoles are tunable and switchable using the interlayer shear displacement. This could explain the ferroelectric-like behavior recently observed in atomically thin $WTe_2$ films[4]. In addition, we reveal that exfoliation of the surface layer flips the out-of-plane spin textures. The presented results will aid in the deeper understanding, manipulation, and further exploration of the physical properties of $WTe_2$ and related atom-layered materials, for applications in electronics and spintronic devices.

**Introduction**

Transition metal dichalcogenides (TMDs) are layered materials that are mostly centrosymmetric at room temperature. $WTe_2$ has garnered widespread interest as an exceptional non-centrosymmetric TMD due to its exotic properties such as an extremely large nonsaturating magnetoresistance (MR), temperature-induced Lifshitz transition, pressure-induced superconductivity, type-II Weyl semimetal, etc.[1–3,5–8] Specifically, it is worth noting that the MR of $WTe_2$ can reach up to $10^5$ % and does not saturate at dozen of tesla[1,5,8]. Although the band structure of $WTe_2$ near the Fermi level ($E_F$) is complicated[8–14], the nonsaturating MR can be adequately explained by the perfectly compensated electron and hole carriers in bulk $WTe_2$ using a two-band model. The two-band model provides an acceptable explanation of the MR, except for several quantum effects, not only in the bulk material, but also in thin film $WTe_2$[5,15,16]. Very recently, electrically tunable spontaneous dipole, i.e., ferroelectric switching, was experimentally discovered in few-layer $WTe_2$[4]. Although $WTe_2$ exhibits thickness-dependent novel properties as described above, quantitative analysis of the electron/hole ratio and its dependence on the thickness as well as the relevant charge distribution from the bulk to monolayer is deficient.[17] Therefore, a systematic theoretical investigation is highly desired.

In this work, we investigate the electronic properties of WTe$_2$ from the bulk to few-layer and monolayer using density functional theory (DFT). We determined that the non-centrosymmetric crystal structure causes vertical charge redistribution in few-layer WTe$_2$, leading to the layer-dependence of the electron/hole carrier ratio, MR, surface dipole, Fermi surface topology, and spin textures. The calculated dipoles reasonably match with those measured experimentally. Moreover, the charge inhomogeneity and the resultant dipole are mechanically tunable and switchable by an in-plane shear displacement of WTe$_2$ layers. Finally, the out-of-plane spin component at the surface is also flippable as a direct result of the exfoliation of an odd number of layers from the existing WTe$_2$ thin layers, which can be explained by the symmetries of even- and odd-layer WTe$_2$ slabs.

**Results and discussion**

The band structure and the charge compensation are calculated for the bulk, monolayer (ML), bilayer (BL), and few-layer WTe$_2$ slabs. These slabs are initially built from the experimental structure of bulk WTe$_2$ (space group *Pmn*2$_1$)[18], as shown in Figure 1a, 1b, and 1c, and then fully relaxed using DFT. The optimized lattice constants are $a$ = 3.49 Å, $b$ = 6.28 Å, and $c$ = 14.24 Å for bulk WTe$_2$, which are close to the experimental values ($a$ = 3.477 Å, $b$ = 6.249 Å, and $c$ = 14.018 Å).[17–19]

Figure 1d displays the isoenergy slices of the band structure of a 6-layer (6L) WTe$_2$ slab. The topological variation of electron and hole pockets is clearly shown by different energy slices near $E_F$. Correspondingly, the electron and hole bands, both have two peaks which cross $E_F$ near the Γ point. These are illustrated in the left panel of Fig. 1e. The densities of the electron ($n$) and hole ($p$) carriers are evaluated from the band-decomposed density of states (DOS) (see the right panel of Figure 1e). Figure 1f and Figure 1g illustrate the calculated carrier density and the $n/p$ ratio, respectively, from ML to 7L WTe$_2$. The carrier densities increase almost linearly as the number of layers increases. ML WTe$_2$ is found to be $p$-type with $n = 1.1 \times 10^{13}$ cm$^{-2}$ < $p = 1.4 \times 10^{13}$ cm$^{-2}$. In contrast, BL WTe$_2$ is slightly $n$-type with $n/p \sim 1.1$. From BL to 7L, WTe$_2$ slabs are all $n$-type with $n/p > 1$. Notably, the $n/p$ ratios of 6L and 7L WTe$_2$ (1.3 and 1.4, respectively) are very close to the experimental values (~ 1.4 and 1.5, respectively)[5]. Generally speaking, $n/p$ gradually increases from ~0.8 (ML) to ~1.2 (3L) and remains close to 1.5 for the thicker slabs. Bulk WTe$_2$ is more electron-rich with $n/p \sim 1.6$, where $n$ and $p$ are $1.2 \times 10^{20}$ and $7.8 \times 10^{19}$ cm$^{-3}$, respectively, corresponding to the areal carrier densities in the $ab$-plane of $1.8 \times 10^{13}$ and $1.1 \times 10^{13}$ cm$^{-2}$, respectively.

In addition to the dependence of $n$ and $p$ on the thickness of WTe$_2$ slab, the electronic structures differ in the top and bottom surfaces. Fig. 2a displays the projected band structure along the ΓX line in the 1$^{st}$ BZ of 6L WTe$_2$. The electron and hole pockets are split into several branches due to the interlayer interaction. The largest surface components appear near the Γ point above $E_F$ and the surface states near 1/3 ΓX, just below $E_F$. The projections of the top and bottom surfaces are apparently different at both positions. The corresponding Fermi surfaces projected on the top and bottom layers, as shown in Figure 2b and Figure 2c, also exhibit distinguishable differences, especially near the Γ point and Fermi arcs[20–22]. Notably, two types of Fermi surfaces in WTe$_2$ were experimentally confirmed and they are attributed to the breaking of the inversion symmetry for the top and bottom Te surface atoms[21,23]. However, their origin remains unknown. It is our opinion that such difference originates from the charge redistribution in WTe$_2$ as a result of the non-centrosymmetric structure, and interlayer interaction, which will be subsequently discussed. Not limited to top and bottom surfaces, other inner layers also have slightly different projections in the band structure and Fermi surfaces, resulting in the layer-dependent $n/p$ in WTe$_2$ slabs.

The layer-decomposed $n/p$ is calculated from the orbital-projected band structure using the projection as the weight for the DOS. As displayed in Figure 2d, the $n/p$ for the top (L1) and bottom (L6) surfaces are significantly different, with one being hole-rich ($n/p \sim 0.6$) while the other is electron-rich ($n/p \sim 1.3$). Meanwhile, both surfaces have a lower $n/p$ than the total $n/p$ (~1.33 for 6L WTe$_2$, Figure 1f), while the layers immediately beneath the surface have a higher $n/p$ than the total value. This indicates that electrons transfer from the surface to the layer below. The $n/p$ difference is more significant in the WTe$_2$ slab whose lattice constants and the surface structure are adopted from experimental measurement by low-energy electron diffraction[19] (we denote this type of slab as LEED structure hereafter). The values for the top/bottom surfaces are 0.4/1.6 and 0.3/1.5 for 6L and 7L WTe$_2$ for the LEED structures, respectively. This is indicative of the sensitivity of $n/p$ with respect to the structure. Therefore, we evaluate the physical properties of both the DFT structure and LEED structure in the following sections. In general, the bottom layers have more electrons and the top layers have more holes.

The layer-dependent carrier compensation leads to different MR in the top and bottom surfaces in WTe$_2$. Within the scheme of the two-band model[5], MR = $\frac{(\alpha\mu_e+\mu_h)^2+\mu_e\mu_h B^2(\alpha\mu_e+p\mu_h)(\alpha\mu_h+\mu_e)}{(\alpha\mu_e+\mu_h)^2+(1-\alpha)^2\mu_e^2\mu_h^2 B^2} - 1$, where $\alpha = n/p$ and $\mu_e$ and $\mu_h$ are the mobilities of the electron and hole carriers, respectively. Since $n/p$ differs for the top and bottom surfaces of WTe$_2$, the MRs of these surfaces (MR$_t$ and MR$_b$) are also different. For example, the MRs of 6L WTe$_2$ with the structure optimized by DFT (hereafter denoted by DFT structure) are demonstrated in Figure 2e. The carrier mobilities of the 6L WTe$_2$ are adopted from experimental values of $\mu_e = 4 \times 10^2$ cm$^2$V$^{-1}$s$^{-1}$ and $\mu_h = 3 \times 10^2$ cm$^2$V$^{-1}$s$^{-1}$.[5] At low magnetic fields ($B$ < 20 T), both MR$_t$ and MR$_b$ increase parabolically and do not exhibit a significant difference. As $B$ increases, the difference becomes larger, which can be seen from the ratio of MR$_b$/MR$_t$ which achieves a value of ~ 1.1 at 50 T and ~ 1.4 at 100 T (Figure 2f). Such a difference is more significant in LEED structures due to the larger difference in $n/p$. For example, MR$_b$/MR$_t$ in 7L WTe$_2$ with a LEED structure is initially ~ 0.7 at $B$ ~ 0 but can reach ~2.2 at B = 50 T (Supplementary Fig. S1).

Herein, we compare our results for $n$, $p$, and $n/p$ with previously calculated values[17]. $n$ and $p$ in ML and BL WTe$_2$ were calculated to be $n = p$ ~ $1.6 \times 10^{13}$ and $1.4 \times 10^{13}$ cm$^{-2}$, respectively[17], which are of the same order of magnitude as our results. Besides, our calculation indicates a larger $n$ ($1.2 \times 10^{20}$ cm$^{-3}$) in bulk compared to the reported value of $7.5 \times 10^{19}$ cm$^{-3}$.[17] These differences may be due to differences in the details of the computational conditions which affects the position of $E_F$, i.e., the exchange-correlation functional, $k$-point grids and the procedure associated with temperature broadening DOS calculations. For the completeness of the discussion, we also examine the dependence of the charge compensation $n/p$ as a function of $E_F$ and temperature (see Supplementary Information).

The charge redistribution results in finite dipole moments $P = \int z\rho(z)dz$ in WTe$_2$ slabs. The variation of $P$ with the number of layers is displayed in Figure 3a. There is a small but finite $P$ of $8 \times 10^{-5}$ (BL) ~ $1.7 \times 10^{-3}$ (6L) eÅ per unit cell for BL WTe$_2$ and beyond. Slabs with an even number of layers have a larger $P$ than those with an odd number of layers. Correspondingly, the bottom surfaces of even-layered WTe$_2$ tend to be more electron-rich than the odd-layered surfaces (Supplementary Fig. S2). We have also calculated $P$ of 6L and 7L WTe$_2$ with the LEED structure and determined these values are approximately 2–3 times the values obtained for the DFT structures (Figure 3a).

Furthermore, $P$ can be mechanically tuned and switched under the interlayer shear displacement $\Delta$, which is defined as the displacement between even and odd layers along the $b$ axis (Figure 3c). This behavior can be explained by the symmetry of the WTe$_2$ slabs, or more specifically, by the fact that adequate interlayer shear displacement is structurally equal to the inversion operation (see Supplementary Information). Figure 3b shows the calculation results for 7L WTe$_2$ as a representative case. In this example, L1, L3, L5, and L7 move $\Delta/2$ along the $+b$ axis, while L2, L4, and L6 move in the opposite direction with $\Delta/2$. When $\Delta$ increases, $P$ decreases almost linearly to 0 at $\Delta \sim 0.2$ Å for the DFT structure and $\Delta \sim 0.4$ Å for the LEED structure, then subsequently flips its direction. The energy barrier to flip $P$ is about 1 (5) meV per unit cell for the DFT (LEED) structure. Correspondingly, there is an apparent charge transfer inside WTe$_2$ as $\Delta$ increases, as shown in Figure 3c for 7L WTe$_2$ with the LEED structure. The most significant charge transfer occurs in the interlayer regions, and the transfer direction is the same for every interlayer region.

The inversion of the total dipole by the interlayer shear displacement in WTe$_2$ as indicated in the preceding section could be realized by surface friction[24], coherent shear phonon generation[25,26], and strain-induced interlayer shear[27]. Notably, the second method is more promising because it is contactless and the shear phonons of the WTe$_2$ are optically pumped[26].

Our calculation results of $P$ reasonably match with that of the experimental data. The experimentally measured $P$ per area for BL WTe$_2$ is about $1 \times 10^4$ e/cm ($1 \times 10^{-4}$ e/Å)[4]. As previously described and shown in Figure 3a, the calculated $P$ beyond 4L WTe$_2$ with DFT structures are approximately $0.9 \times 10^{-3} \sim 1.7 \times 10^{-3}$ eÅ *per unit cell*, corresponding to $0.4 \times 10^{-4} \sim 0.8 \times 10^{-4}$ e/Å dipoles per area, considering the area of the unit cell as approximately 22 Å². The dipoles per area of 6L and 7L WTe$_2$ with the LEED structures are $1.5 \times 10^{-4}$ and $1.4 \times 10^{-4}$ e/Å, respectively. Therefore, the dipole values calculated for both the DFT and LEED structures are reasonably of the same order of magnitude as the experimental results.

The presence of the surface dipole and the resulting electron/hole rich characteristic of the top and bottom layers explain the difference in the Fermi surfaces projected on the top and bottom layers. Since the direction and magnitude of the surface dipole are determined by the non-centrosymmetric structure of WTe$_2$, the exfoliation of the surface layer does not significantly change the characteristic of the Fermi surface for the top and bottom layers (Supplementary Fig. S3). However, the exfoliation alters the spin texture. We determined that exfoliation of the surface layer in WTe$_2$ switches the sign of the out-of-plane spin ($S_z$) texture, while the in-plane spin texture ($S_x$ and $S_y$) are essentially unchanged, as demonstrated in Figure 4 for 7L WTe$_2$ and the top-surface exfoliated 7L WTe$_2$. The behavior of the spin textures can be explained by the transformation of the spin operators under symmetric operations in the WTe$_2$ slabs (see Supplementary Information).

**Method**

DFT calculations were performed using the generalized gradient approximation (GGA) as implemented in the Vienna Ab-Initio Simulation Package (VASP).[28,29] The projector augmented wave pseudopotentials are employed.[28] The optB86b-vdW exchange-correlation functional is adopted to take into account van der Waals (vdW) interactions.[30–33] An energy cutoff of 400 eV and Monkhorst–Pack grids of 15×9×1 *k*- points is utilized.[34] The force tolerance is set to $10^{-4}$ eV/Å for geometry optimization, and the energy tolerance is set to $10^{-8}$ eV for self-consistent calculations. We included both the van-der-Waals correction and the spin-orbit coupling (SOC) using the method implemented in VASP by Hyun-Jung Kim, *et al.*[35] A vacuum distance in excess of 20 Å is utilized for slab calculations, and a dipole correction is employed to avoid interaction between spurious periodic images. The total and layer-projected electron/hole carrier ratios in this work are calculated using a $100 \times 100 \times 1$ ($60 \times 60 \times 20$ for bulk) *k*-point grid by the integration of band-decomposed projected density of states modulated by the Fermi distribution. For the band-decomposed density of states (DOS) calculations, the Dirac δ function is approximated by the Gaussian function with the width σ corresponding to temperature broadening.

**Acknowledgement**

The following financial supports are acknowledged: Grants-in-Aid for Scientific Research on Innovative Areas (MEXT KAKENHI Grants No. JP26102017, JSPS KAKENHI Grants No. JP15H03561, No. JP17H05215, World Premier International Research Center Initiative (WPI) The calculations were performed by using the computer facilities of the Institute of Solid State Physics (ISSP Super Computer Center, University of Tokyo), and RIKEN (HOKUSAI GreatWave).

**Author contributions**

Z.N. performed the calculations and the data analysis. Z.N. and E.M. wrote the manuscript with crucial inputs from R.A., C.L.L., N.T., and S.W. The LEED structure data is provided by K.K. and N.T. This project is supervised by E.M. and S.W.

**Competing financial interests**

The authors declare no competing financial interests.

**Figures and captions**

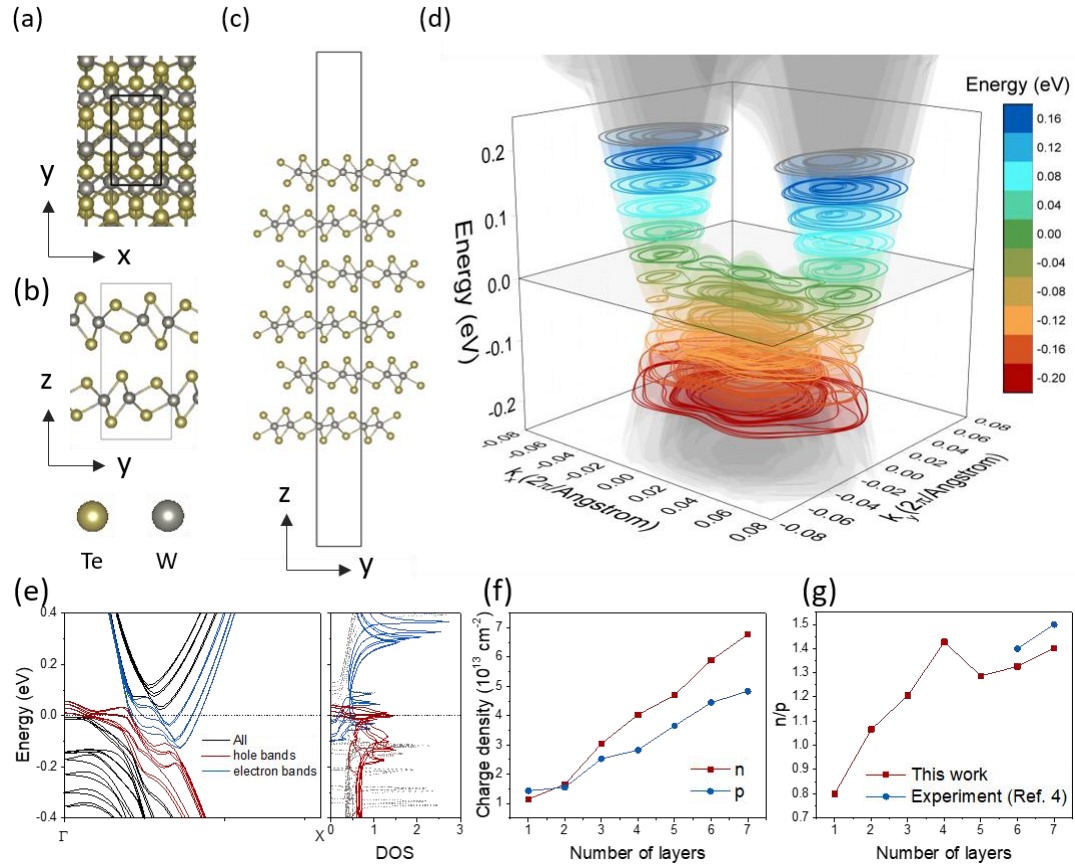

**Figure 1| Geometry, band structure and carrier compensation of WTe$_2$.** (a) Top and (b) side view of bulk WTe$_2$; (c) Side view of 6L WTe$_2$. (d) Band structure with SOC near the Γ point of 6L WTe$_2$. The horizontal axes in (d) are $k_x$ and $k_y$, while the vertical axes indicate the energy of the bands. The Fermi level is aligned to zero. The electron and hole pockets are clearly shown at the Fermi surface (deep green circles); (e) Schematic of electron and hole bands along ΓX in 6L WTe$_2$ (left) and the corresponding band decomposed DOS used in the calculation of charge compensation (right). (f) Total carrier densities per area of electron and hole in WTe$_2$ from ML to 7L; (g) Ratio between electron

carriers and hole carriers ($n/p$). Blue dots are experimental values taken from Ref. 5. In the calculations of (f) and (g), the temperature is set at 1.4 K, which is close to the low temperatures used in previous experiments.[5]

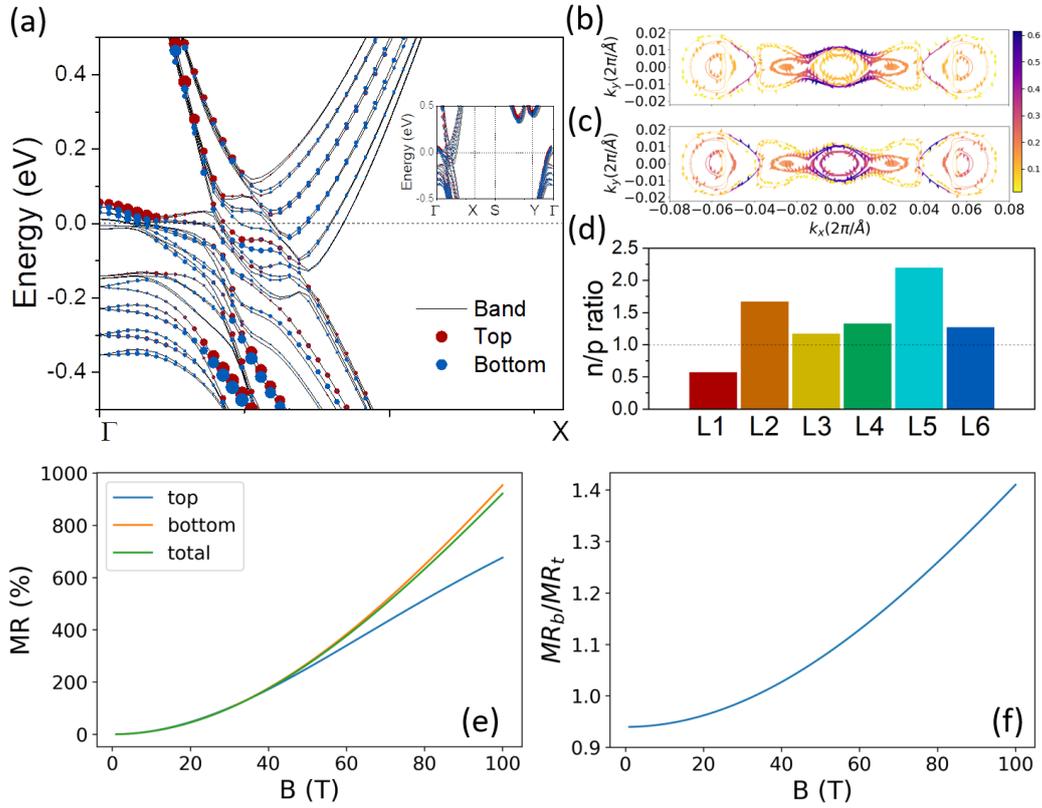

**Figure 2| Differences between top and bottom surfaces in 6L WTe$_2$.** (a) Band structure of 6L WTe$_2$ along ΓX. Red and blue dots are scaled to top and bottom surface projections, respectively. The band structure along the whole high symmetric line is shown in the inset; (b) Projected $n/p$ for each layer; (c & d) Fermi surfaces and in-plane spin textures for the (c) top (1L) and (d) bottom (6L) surfaces of 6L WTe$_2$. The color bar denotes the projection of the corresponding surface, while the tiny arrows represent the direction of the in-plane spin ($S_x$, $S_y$). (e) The surface and total MR of 6L WTe$_2$ estimated from the two-band model. (f) Ratio between the MR of the bottom (MR$_b$) and top (MR$_t$) surfaces in 6L WTe$_2$.

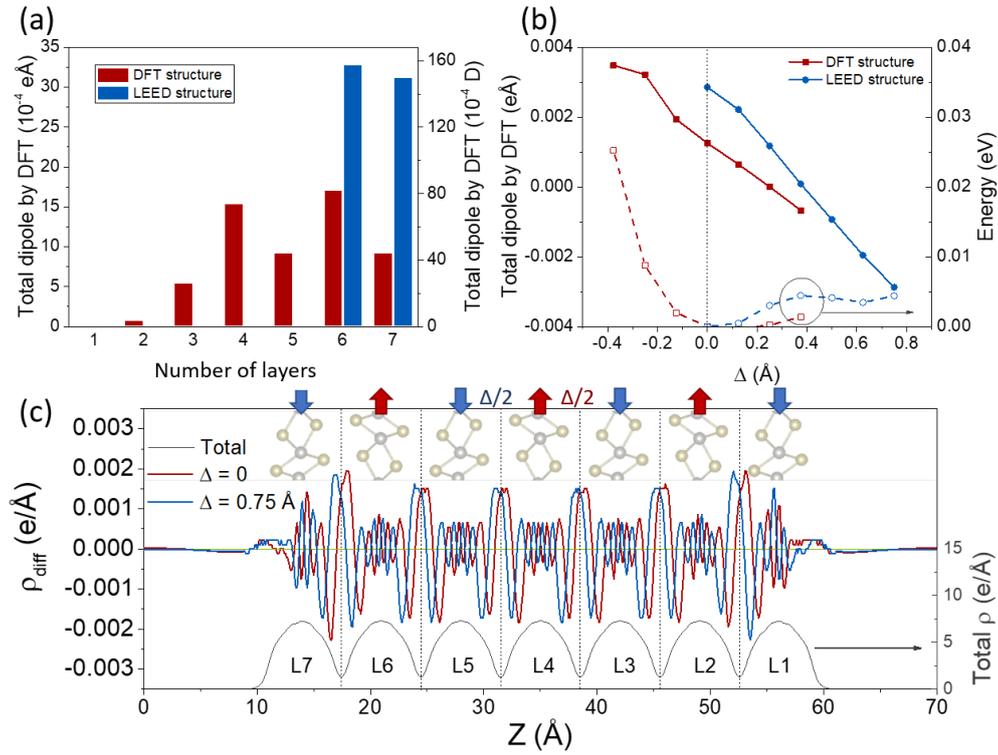

**Figure 3| Switchable dipole in few-layer WTe$_2$.** (a) Intrinsic dipole of WTe$_2$ with different thicknesses. In addition to the dipole of the DFT optimized structures, those with geometries measured by LEED are also given for 6L and 7L cases. (b) Total dipole (solid lines) and energy (dashed lines) of 7L WTe$_2$ under interlayer shear displacement Δ. (c) Total and differential vertical charge density ($\rho$, $\rho_{\text{diff}}$) of 7L WTe$_2$ (LEED structure) with different Δ. The displacement schematic is presented in the top of the figure by blue and red arrows. The reference charge density is at Δ ~ 0.4 Å, which has almost zero dipole.

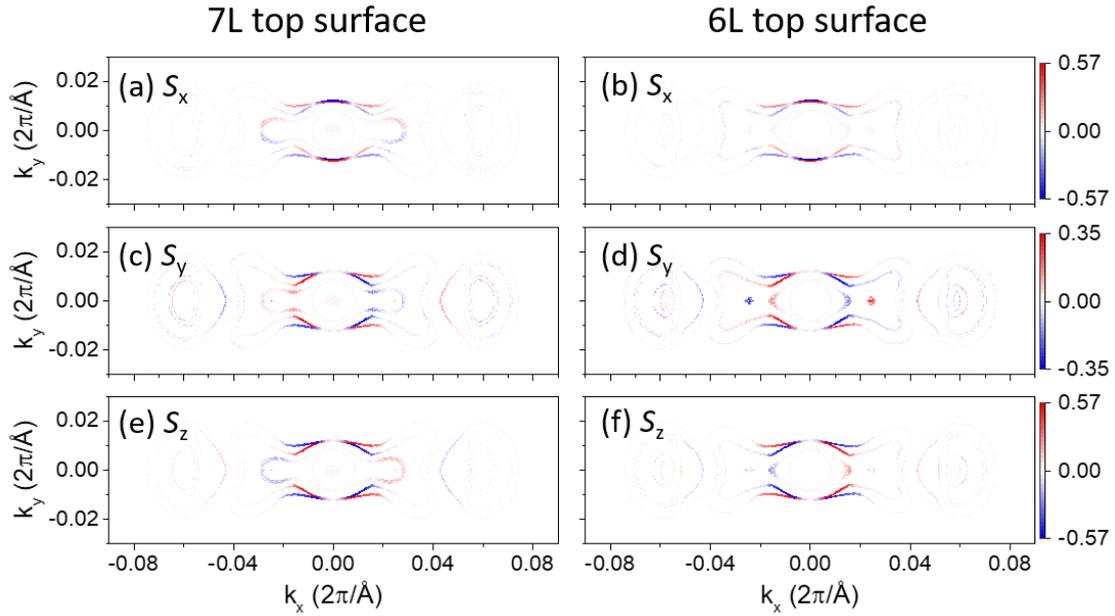

**Figure 4| Spin texture under surface exfoliation.** (a, c, e) Top surface spin texture of 7L WTe$_2$ of the LEED structure; (b, d, f). The new top surface spin texture after exfoliation of the old top surface. The new structure becomes 6L WTe$_2$, and surface relaxation measured from LEED is adopted. The sign of $S_x$ and $S_y$ persists, while that of $S_z$ flips.

# Mechanically tunable spontaneous vertical charge inhomogeneity in few-layer WTe$_2$

Zeyuan Ni[1,*], Emi Minamitani[1], Kazuaki Kawahara[2], Ryuichi Arafune[3], Chun-Liang Lin[2], Noriaki Takagi[2], Satoshi Watanabe[1]

[1]Department of Materials Engineering, The University of Tokyo, Tokyo 113-8656, Japan

[2]Department of Advanced Materials Science, The University of Tokyo, 5-1-5 Kashiwanoha, Kashiwa, Chiba 277-8561, Japan

[3]International Center for Materials Nanoarchitectonics (WPI-MANA), National Institute for Materials Science, 1-1 Namiki, Tsukuba, Ibaraki 305-0044, Japan

**Dependence on the calculation conditions**

Previously we have confirmed that DFT can reproduce the fine trend of surface relaxation in WTe$_2$ measured by low-energy electron diffraction (LEED).[1] In addition, band structures of all cases show clear electron and hole pockets in good agreement with previous literature.[2] The switchable dipole is confirmed by using different computational settings, including using/excluding SOC, changing functionals, pseudopotentials, k-points, and vdW correction methods (Figure S4 and Figure S5).

**Dependence of n/p on Fermi level shifting and virtual temperature.**

The charge compensation calculated for bulk and slab WTe$_2$ are extremely sensitive to the Fermi level shifting $\Delta E_f$, changeing one order of magnitude for merely $\Delta E_f \sim$ 30-50 meV in all cases. In addition, all cases can achieve exactly $n/p = 1$ within $\Delta E_f < 10$ meV (Figure S6). The charge compensation is also affected by the temperature broadening but not as significant. When the virtual temperature increases from 1.4 K to 300 K, $n/p$ in all cases decreases to 0.7~0.9 (Figure S7).

**Mechanism and explanation of switchable P**

The switchability of $P$ could be explained by the symmetry of WTe$_2$ slabs. Below we will take the 7L WTe$_2$ of LEED structure as the example. At $\Delta = 0$, 7L WTe$_2$ has the symmetry of $P1m1$ and is non-centrosymmetric, allowing for the presence of finite $P$. At $\Delta \sim 0.4$ Å, its symmetry becomes $P12_1/m1$ and obtains inversion symmetry, so $P$ must be 0. At $\Delta \sim 0.8$ Å, it recovers the $P1m1$ symmetry, but the structure is almost identical as the inversed intrinsic case ($\Delta = 0$), i.e. the 7L WTe$_2$ slab is "flipped" by the interlayer shear displacement. Naturally, the direction of $P$ is switched, while the magnitude of $P$ keeps similar. Additionally, if only the top surface is moved by $\Delta$, $P$ will decrease about half the amount of that induced by the layer shear displacement (Figure S8); if the mid-most layer is moved, $P$ hardly changes (Figure S8). These facts demonstrate that $P$ originates from the interlayer interaction and non-centrosymmetric structure of WTe$_2$, and the outermost 1 or 2 surface layers play an important role to determine the value of $P$, due to the screening effect from WTe$_2$'s semimetallic nature. In conclusion, the dipole mainly comes from the symmetry of WTe$_2$ and mostly localized at the surface and the layer beneath.

**Explanation on the behavior of spin textures when the surface layer is exfoliated**

Take the out-of-plane spin texture $S_z(k_x, k_y)$ as an example, WTe$_2$ slabs have $P1m1$ symmetry with a mirror plane perpendicular to the $x$ axis, so $S_z(k_x, k_y) = -S_z(-k_x, k_y)$. The time reversal symmetry rules that $S_z(k_x, k_y) = -S_z(-k_x, -k_y)$. Considering also the $Pmn2_1$ symmetry in bulk WTe$_2$ crystal, the L1 ~ L6 part of 7L WTe$_2$ can be transformed to the L2 ~ L7 part (i.e. top-surface exfoliated 7L WTe$_2$) by a mirror plane perpendicular to the $y$ axis, so the top-surface spin-texture after exfoliating the top-layer from the 7L slab should fulfill the following constraint: $S_{z\_7L}(k_x, k_y) = -S_{z\_7L\_top\_exfoliated}(k_x, -k_y)$. By applying the three symmetric requirements, one will get the relations of $S_{z\_7L}(k_x, k_y) = S_{z\_7L}(k_x, -k_y) = -S_{z\_7L\_top\_exfoliated}(k_x, k_y)$, so the switch of the sign in $S_z$ by surface exfoliation is explained. The unchanged sign of $S_x$ and $S_y$ can be explained similarly.

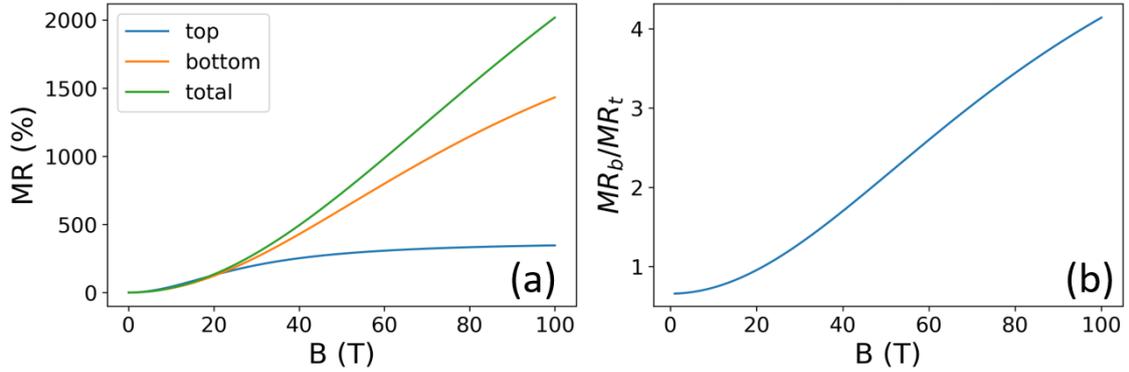

Figure S 1: (a) MR and (b) MR ratio between bottom and top surfaces of 7L WTe$_2$ with LEED structure at different magnetic field.

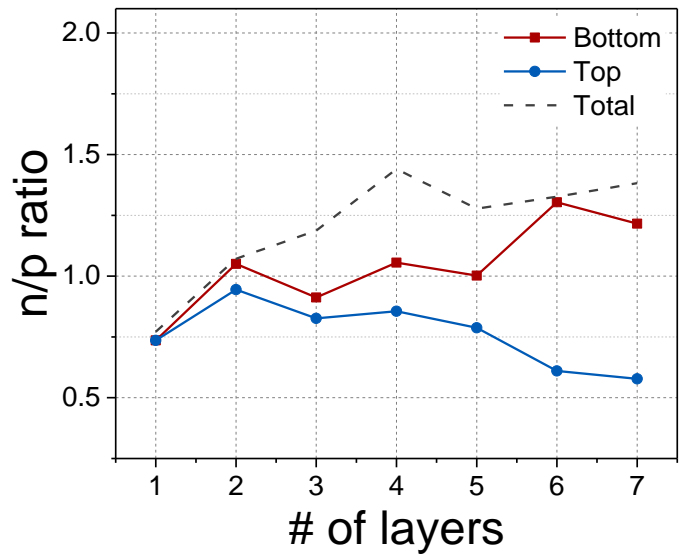

Figure S2: Charge compensation of top and bottom surfaces in WTe$_2$ slabs.

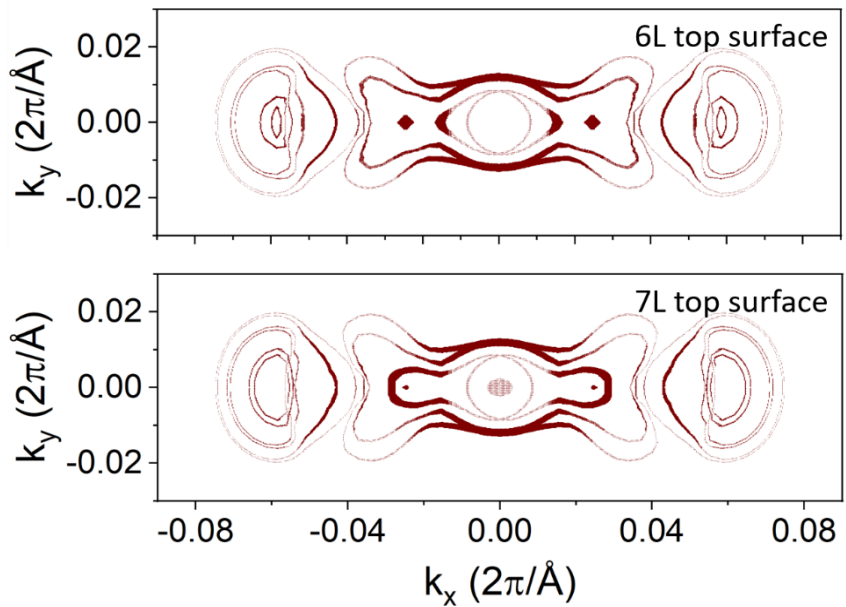

Figure S3: Comparison of the Fermi surfaces of the top surfaces between 6L (=7-1L by top surface exfoliation) and 7L WTe$_2$ with LEED structures. The size of dots is proportional to the surface projections. Their shapes are similar.

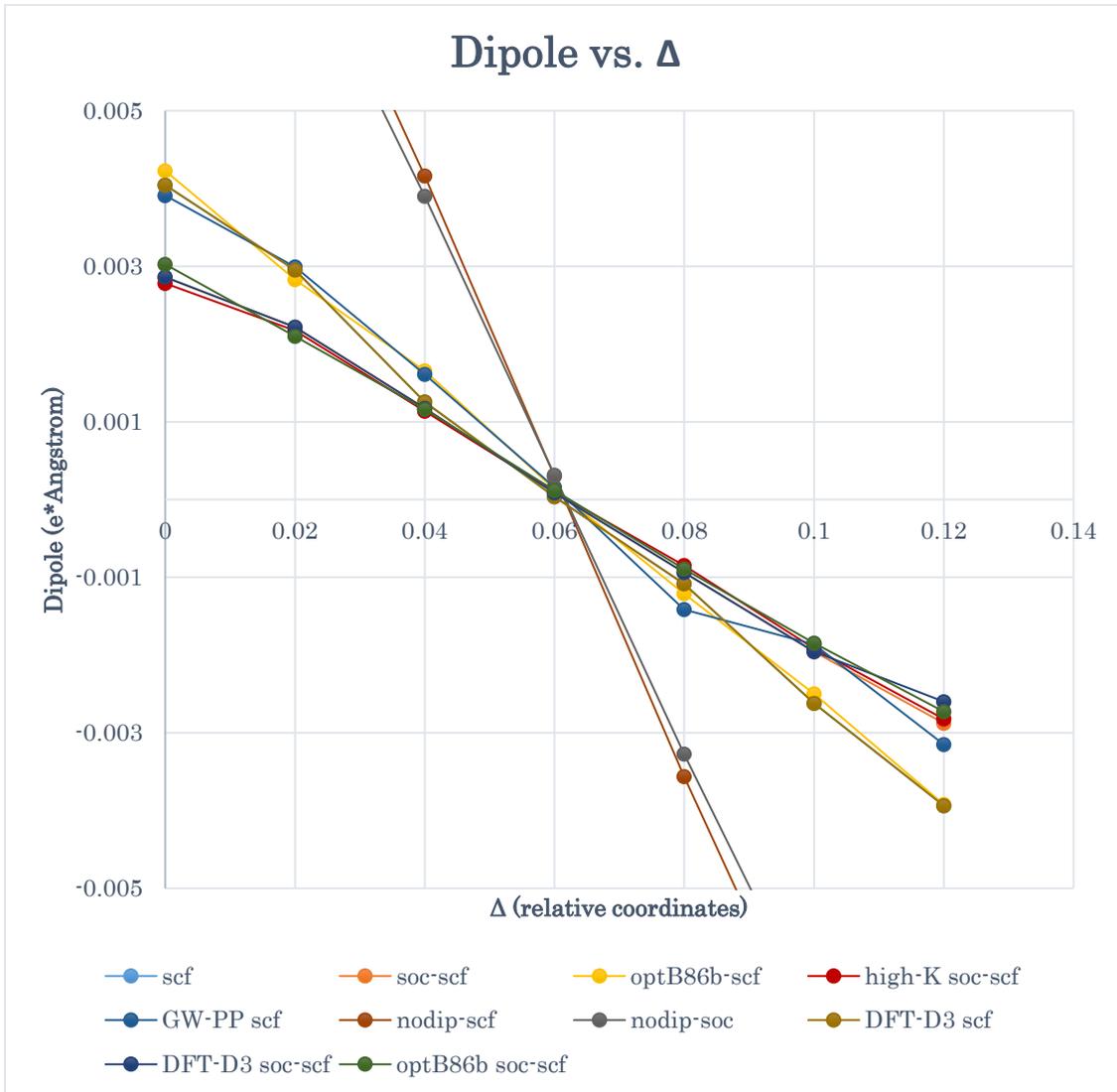

Figure S4: Dipole of 7L WTe$_2$ with LEED structure under interlayer shear displacement using different computational parameters. The meaning of legends are explained in the following table.

Table S1: Meaning of legends in Figure S4 and S5. *used in manuscript.

| Label | scf | soc-scf* | optB86b* | DFT-D3 | GW-PP | no-dip | High-K |
|---|---|---|---|---|---|---|---|
| **Meaning** | w/o SOC | full SCF-SOC | type of vdW correction | | using pseudopotential generated for GW calculations instead of default ones | Excluding dipole corrections | 30x18x1 $k$-points in SCF |

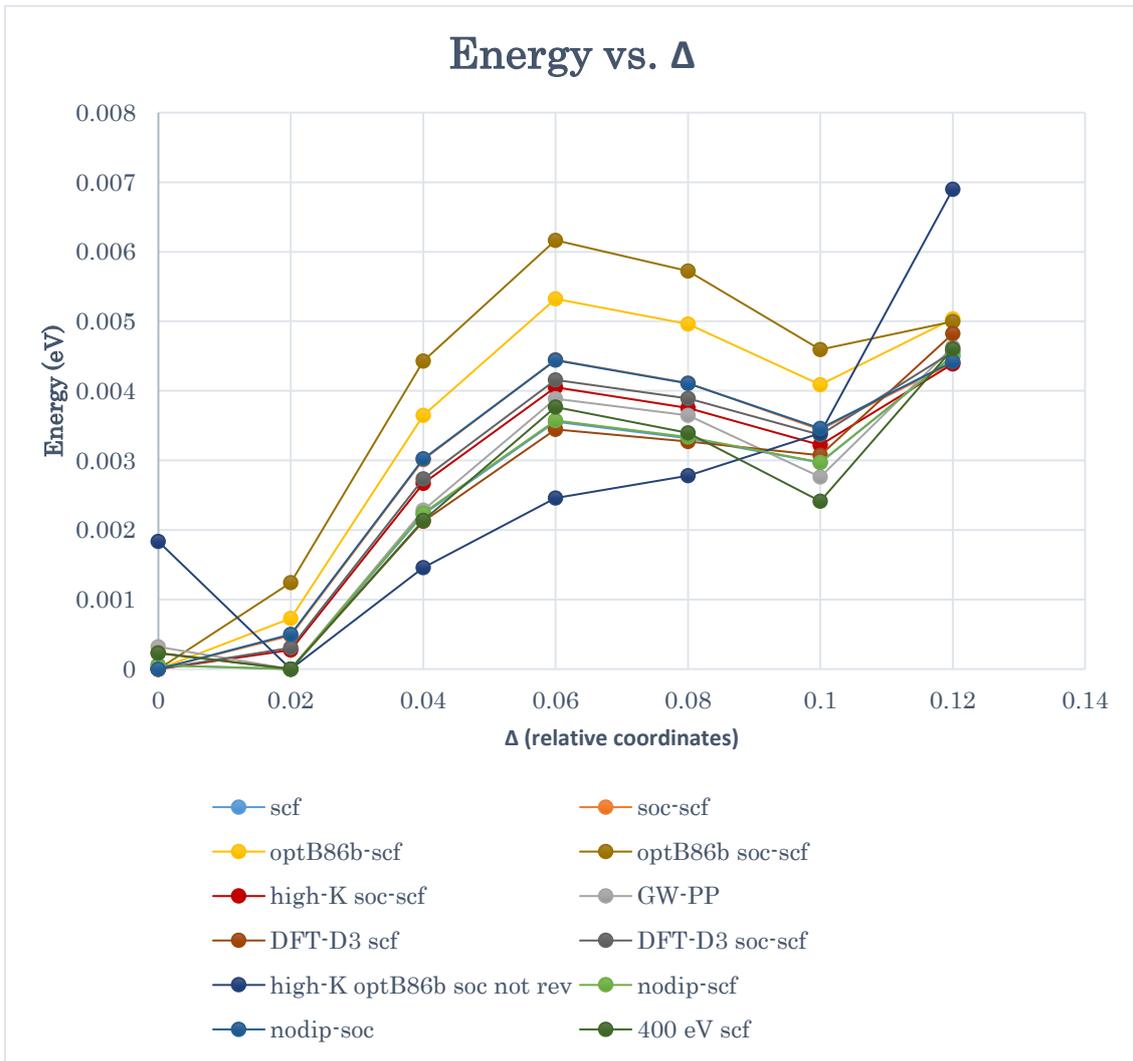

Figure S5: Energy of 7L WTe$_2$ with LEED structure under interlayer shear displacement using different computational parameters.

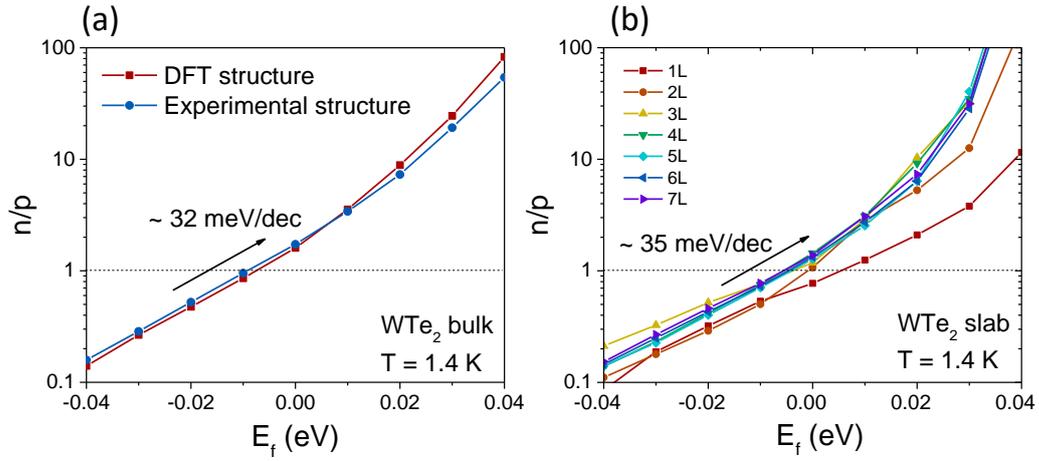

Figure S6: Charge compensation *n/p* in (a) bulk and (b) slab WTe$_2$ with different Fermi level shifting. The slopes are around 30 ~ 50 meV/dec at the original Fermi level $E_f$, indicating that a shift of 30 ~ 50 meV in $E_f$ will result in a change of one order of magnitude in *n/p*. The experimental structure is from Ref. [3]. All results are DFT calculations.

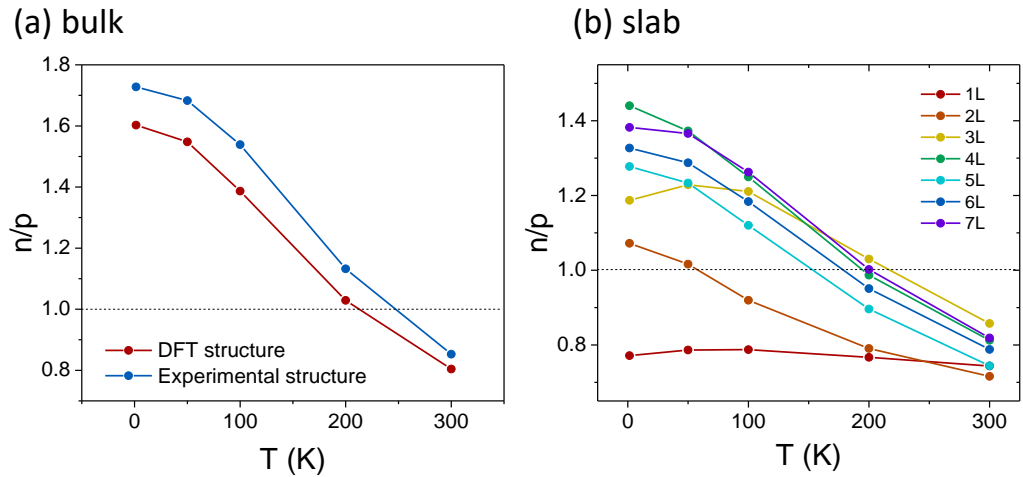

Figure S7: Temperature dependency of n/p in (a) bulk and (b) slab WTe$_2$. The experimental structure is from Ref. [3]. All results are DFT calculations.

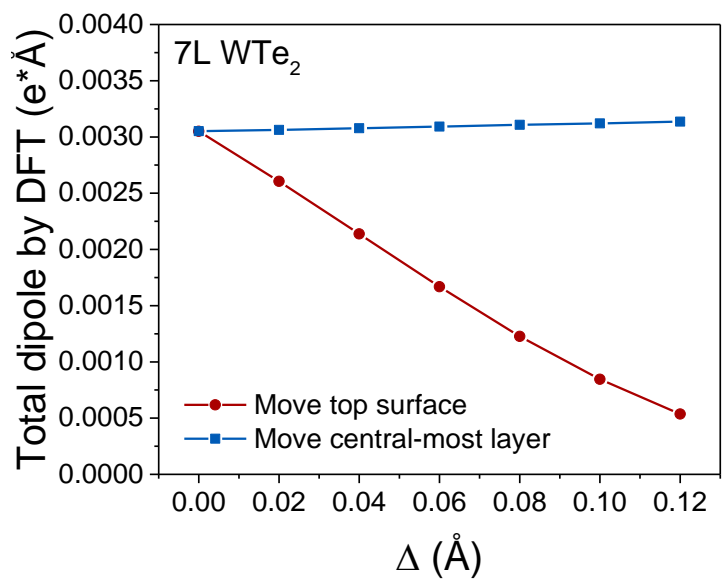

Figure S8: Dipole of 7L WTe$_2$ with LEED structure (see manuscript for definition) under the displacement of top surface and central-most layer, respectively.